\documentclass{elsart}
\usepackage{graphicx}
\usepackage{amssymb}
\newcommand{\bra}[1]{\, \langle\,{#1} \,|\,}
\newcommand{\ket}[1]{\,| \, {#1} \,\rangle  \,}

\newcommand{\tgh} {{\rm tgh} \,}

\newcommand{\MyRe} {{\mathcal R} e \,}
\newcommand{\MyIm} {{\mathcal I} m \,} 

\begin{document}

\begin{frontmatter}

\title{Modulation of dephasing due to a spin-boson 
environment\thanksref{label1}} 
\thanks[label1]{
The authors are pleased to dedicate this paper to Uli Weiss 60th
birthday}

\author[1]{E. Paladino}
\ead{epaladino@dmfci.unict.it}
\author[2]{, M. Sassetti}
\author[1]{and G. Falci}

\address[1]{ Dipartimento di Metodologie Fisiche e Chimiche 
(DMFCI), Universit\`a di Catania,
viale A. Doria 6, 95125 Catania, Italy \& MATIS-Istituto Nazionale per la 
Fisica della Materia, Catania, Italy}
\address[2]{Dipartimento di Fisica, INFM-Lamia, Universit\`a di Genova, Via 
Dodecaneso 33, 16146 Genova, Italy}
\begin{abstract}
We study the reduced dynamics of a spin (qubit) coupled to a spin-boson 
environment
in the case of pure dephasing.
We derive formal exact expressions which 
can be cast in terms of exact
integro-differential master equations.
We present results for a SB environment with ohmic dissipation at finite 
temperatures. For the special value of the ohmic damping
strength $K=1/2$ the reduced dynamics is found in analytic form.
For $K \ll 1$ we discuss the possibility of modulating the effect of 
the SB environment on the qubit. In particular we 
study the effect of the crossover to a slow environment dynamics, which 
may be triggered by changing both 
the temperature and the system-environment coupling. 
\end{abstract}


\end{frontmatter}
\section{Introduction}
\label{intro}
The possibility of controlling the dynamics of a complex quantum
system would open a wide scenario for both fundamental and applied physics.
New physical properties emerging at the nanoscale may yield new paradigms for
nanoelectronics, an example being quantum information 
processing~\cite{kn:qcomp}.
The basic operations of a quantum logic device can be described in 
its {\em computational space} defined by a set of 
degrees of freedom which can be prepared (write) and measured (read)
and of course evolve coherently. 
However the  Hilbert space of the device is
much larger than the computational space, and effects of the additional 
degrees of freedom may be relevant for the reduced dynamics we aim to 
control~\cite{kn:zurek,kn:palma96}.
These environmental degrees of freedom may be ``internal'' to the device,
which is a many-body object, or ``external'', belonging to the 
read/write circuitry or to auxiliary devices (buses or other idle or working 
qubits). They determine loss of coherence, which is ultimately due to the 
entanglement between system and 
environment~\cite{kn:zurek}.

Much of the effort in the analysis of decoherence in quantum logic devices 
has been focused on 
model environments made of many weakly coupled harmonic 
oscillators~\cite{kn:zurek,kn:rmp}. The 
key feature of these linear models is that the effects on the 
reduced dynamics of the system can be described
in terms of simple statistical information on the environment, spectral 
densities or power spectra, even if a detailed knowledge of its
microscopic structure is not known~\cite{book,Leggett,kn:dissipative}.
However a slightly more detailed knowledge of the environment dynamics 
is needed in certain situations, for instance if the environment has memory
on the time scale of the system dynamics~\cite{kn:cohen,PRL}. 

Bistable impurities located 
close to a solid state device which produce $1/f$ noise~\cite{kn:weissman} 
are a physical realization of an environment with memory. 
They have been recognized as the major source of decoherence in 
recent experiments~\cite{kn:exp} 
demonstrating quantum coherent dynamics in superconducting 
devices. Such an environment has been modeled in Ref.~\cite{PRL}, 
with a set of Fano impurities~\cite{mahan} which randomly 
polarize the device. 
A peculiar effect of this environment, related to its non linear 
nature, is that the reduced qubit dynamics shows various qualitatively 
different regimes depending on the switching rate of the impurities.  
In particular dephasing due to slow impurities is not described by 
a single rate and memory effects of the environment enter the qubit
dynamics.

The main goal of this work is to discuss a situation where 
the reduced dynamics of the system is able to probe 
different regimes of the environment dynamics. 
To this end we introduce a model which exhibits tunable crossover
effects. 
We take as the environment an impurity described by a spin-boson (SB) 
model~\cite{book,Leggett}. The Hamiltonian 
is introduced in section \ref{sec:model}, where we also discuss  
applications to physical problems. In section \ref{sec:formal} the formal 
solution for the off-diagonal element $\rho_{01}(t)$ 
of the RDM is found, in the limit of
pure dephasing and ohmic damping in the scaling regime. In section 
\ref{sec:gen-master} a generalized master equation for $\rho_{01}(t)$ is
derived and the analytic structure of the kernels is exploited in section
\ref{sec:reduced-dynamics} to study the character of the behavior of 
$\rho_{01}(t)$. Finally in section \ref{sec:conclusions}
we sketch a phase diagram for this problem.

\section{Model}
\label{sec:model}
We study the following Hamiltonian ($\hbar=1$)  
\begin{eqnarray}
\label{Hamiltonian}
H &=& 
- \, \frac{\epsilon}{2} \, \sigma_z  -
 \frac{E_J}{2} \,\sigma_x  - \frac{v}{2} 
 \, \tau_z \, \sigma_z + H_{SB} \, ,
\\
H_{SB} &=& \frac{\epsilon_{SB}}{2} \, \tau_z 
         -  \frac{\Delta}{2} \,\tau_x 
        -  \frac{1}{2}\tau_z  \, X
        +  \sum_\alpha  \omega_\alpha  a^\dagger_\alpha a_\alpha 
\quad ; \;
 X=\sum_{\alpha} \lambda_\alpha 
                        (a^\dagger_\alpha + a_\alpha) \, .
\quad
\nonumber
\end{eqnarray}
It describes two spins ($\vec{\sigma}$ and $\vec{\tau}$) coupled to each 
other by the interaction $v$. 
The spin  $\vec{\tau}$ is also coupled to a 
collective coordinate $X$ of a set of otherwise 
noninteracting harmonic oscillators 
(operators $a^\dagger_\alpha$ and $a_\alpha$). The Hamiltonian $H_{SB}$ 
represents a spin-boson model~\cite{book,Leggett}, which describes a 
two state system with tunneling amplitude $\Delta$ and bias $\epsilon_{SB}$
coupled to a set of harmonic oscillators.
This set of bosons is characterized by the
spectral density of the operator  $X$, given by
$G(\omega) = \sum_\alpha \lambda{}_\alpha^2 \delta(\omega - \omega_\alpha)$.
The Hamiltonian Eq.(\ref{Hamiltonian}) is very rich and may model many 
different physical problems.  
In this work we will regard $\vec{\sigma}$ as describing a qubit 
coupled with an environment consisting of a bistable impurity 
($\vec{\tau}$) described by the SB model, $H_{SB}$.
We are primarily interested in the reduced dynamics of 
the qubit, resulting from tracing out 
all the environmental degrees of freedom, i.e. the 
harmonic modes {\em and} the spin $\vec{\tau}$. 
Physically $\vec{\sigma}$ may represent a Josephson quantum logic
device where external voltage gates and magnetic fields are used to 
tune the effective charging energy $\epsilon$ and Josephson 
coupling $E_J$~\cite{kn:rmp,kn:josephson-qubit-theory}. 
The environment may be a charge close to the 
device undergoing phonon-assisted tunneling between two positions.
We will consider an 
ohmic spectral density $G(\omega)$~\cite{book}, i.e.
we choose a set of free bosons such that 
the power spectrum of the collective coordinate $X$ is 
given by  
\begin{eqnarray}
\label{eq:powerspec-bosons}
S_X (\omega) \,&=&\, \int_{-\infty}^{\infty} 
\, dt \, \frac{1}{2} \,
\langle X(t) X(0) + X(0)X(t) \rangle \, e^{i \omega t} \, =  
\nonumber\\
&=& \,
\pi \, G(\omega) \, \coth {\beta |\omega| \over 2} \,=\, 
2\pi \, K \, |\omega| \,  e^{- |\omega|/ \omega_c}
\, \coth {\beta |\omega| \over 2} \, ,
\end{eqnarray}
where $\omega_c$ represents the high frequency cut-off of the harmonic
modes.
The case of ohmic spectral density is interesting because the SB 
model has a rich phase diagram resulting in an environment where
memory effects may be tuned by the temperature $T$. 
We will present results for 
$K \ll 1$ focusing on nonzero temperatures, 
$T > \Delta_r = \Delta (\Delta/\omega_c)^{K/(1-K)}$, where
$\Delta_r$  is the 
renormalized tunneling amplitude in $H_{SB}$~\cite{book}. Physically 
this regime describes a SB impurity active in producing decoherence 
for the qubit $\vec{\sigma}$. From a technical point of view 
this regime requires some care. 
For instance the dephasing rate for 
$\langle \tau_z \rangle$ using the standard secular master 
equation~\cite{kn:cohen,book} in the couplings $\lambda_\alpha$ 
(for $\varepsilon_{SB} = v =0$) reads ${S_X (\Delta_r)/4}$~\cite{book}.  
For  $T > \Delta_r$ it reduces to 
\begin{equation}\label{eq:eff-coupling}
\gamma(T) \,=\,  \pi \, K \, k_B T  \, , 
\end{equation}
this result being valid only if $\gamma(T) \ll \Delta_r$~\cite{kn:cohen,book}.
Thus for large enough temperatures an analysis which goes beyond the standard 
secular master equation is needed~\cite{GME,book}.

We mention finally other physical 
problems where the model we study may be relevant. First of all
the Hamiltonian Eq.(\ref{Hamiltonian}) may describe a system of two-qubits 
($\vec{\sigma}$ and $\vec{\tau}$) subject to gaussian noise, the ohmic case 
describing the effect of a resistive circuitry. The low temperature regime,
which may describe a working two-qubit gate, has been investigated using
the standard secular master 
equation~\cite{kn:grifoni-governale}. Instead for temperatures $T > \Delta_r$ 
our model may describe a working qubit ($\vec{\sigma}$) coupled with a idle
qubit ($\vec{\tau}$) not perfectly switched off. Finally the spin $\sigma$ may
represent a measuring device~\cite{kn:devoret-schoelkopf} 
for the mesoscopic system described by $H_{SB}$. 
This picture is an alternative framework for our work, namely the 
reduced dynamics of $\vec{\sigma}$ provides the result of the measurement 
on a SB system. If the coupling $v$ is weak and the dynamics of the SB is fast
enough only 
the power spectrum of the coupling operator $v \tau_z$
\begin{equation}
\label{eq:powerspec}
S_\tau (\omega) \,=\, \int_{-\infty}^\infty \hskip-3pt dt \;
\frac{v^2}{2} \, 
\bigl(\langle \tau_z(t)   \tau_z(0) + \tau_z(0) \tau_z(t) \rangle 
- \langle \tau_z \rangle_\infty^2 \bigr)
\;
\mathrm{e}^{i \omega t}
\end{equation}
enters the dynamics of $\vec{\sigma}$, 
whereas if the SB has a slow dynamics the spin $\vec{\sigma}$ is able
to detect also details of the dynamics of $\tau$ which go beyond 
$S_\tau (\omega)$.

\section{Formal solution for the reduced density matrix}
\label{sec:formal}

We consider the pure dephasing regime $E_J=0$ where 
$[ {H}, \sigma_z ] =0$, so no relaxation process takes place and
dephasing is due to fluctuations of the energy difference of 
system, i.e. $\sigma_z$,  eigenstates.
Hence the diagonal elements of the reduced density matrix of the
qubit in the $\sigma_z$ basis (populations) are constant, whereas
off-diagonal elements (coherences) decay.
All the information on the reduced dynamics is  included in the 
coherences. In this section we derive an exact formal expression for
the coherence $\rho_{01}(t) \equiv \bra 0 \rho(t) \ket 1$.

We choose a factorized initial density matrix
$W(0)=  \rho(0) \otimes w_{SB}(0)$, with the spin-boson environment 
in the general state  $w_{SB}(0)$.
Product initial states are relevant in quantum computing, one of the
main requirements being the
possibility to prepare the qubit in well defined initial states.
The coherence $\rho_{01}(t)$ is thus related to the following
correlation function involving the spin-boson variables
\begin{equation}
\rho_{01}(t) = \rho_{01}(0) \, e^{i \epsilon t} \, 
{\rm Tr}_{SB} \big\{ e^{-i H_{-} t} \, 
w_{SB}(0)
e^{i H_{+}t} \big\}\, \equiv \rho_{01}(0) \, e^{i \epsilon t} \, C_{-+}(t)
\label{element}
\end{equation}
where $H_{\pm}= H_{\rm SB} \pm \frac{v}{2} \tau_z$.
We evaluate the correlator $C_{-+}(t)$ using a functional integral 
approach. 
In the derivation we consider a dimensionless continuous ``coordinate'' $q$, 
and specify later to the spin states ($\vec{\tau}$).
The trace over the environmental degrees of freedom in the 
coordinate basis $\ket{q} \ket{\{ x_\alpha \} }$ reads
\begin{eqnarray}
C_{-+}(t) = \int dq{}_f \int dq{}_i  dq{}_i'
&& \int dx_f  \, dx_i \, dx{}_i' \,
 \;\;  K_{-}(q_f, x_f, t; q_i, x_i, 0) \nonumber \\
&&   
\!\!\!\!\!\bra{q_i,  x_i} \, w_{SB}(0) \,\ket{q_i', x_i'} \;
K_{+}^*(q_f, x_f, t;  q'_i, x'_i, 0) 
\label{c_pm1}
\end{eqnarray}
where we have introduced the compact notation 
\begin{equation}
\quad 
x \equiv \{ x_\alpha \}
\;,\qquad \qquad
\int dx  \equiv \prod_{\alpha} \int dx_\alpha \, .
\end{equation}
The Feynman propagators $K_{\pm}$ are associated to the two Hamiltonians
$H_{\pm}$ and can be  expressed as a real-time path integral 
$$
K_{\pm}(q_f, x_f, t; q_i, x_i, 0) =  
\int\! {  D} q {  D} x \exp \left [ \,  i ({  S}^{}_{\pm} [q] 
+
{  S}^{}_{\rm{B}} [x] +  {  S}^{}_{\rm{SB}} [q,x] )
              \, \right ] \, ,
$$
with the constraints $q(0) = q_i$, $q(t)= q_f$
and $x(0) = x_i$, $x(t) = x_f\,$. ${  S}^{}_{\rm{B}} [x]$ 
and ${S}^{}_{\rm{SB}} [q,x]$ are the actions corresponding to the
free boson and the spin-boson interaction terms
in (\ref{Hamiltonian}) respectively.
We now specify to an initial non equilibrium
state of the factorized form
$w_{SB}(0) =  w_\tau(0) \otimes w_\beta$, where the
harmonic oscillators start from the thermal equilibrium state
$w_\beta$, and $w_\tau(0)$ is a general density matrix for the
system of coordinate $q$.
Then $C_{-+}(t)$ reads
\begin{eqnarray}
&&C_{-+}(t) = \int dq{}_f \int dq{}_i  dq{}_i' \,
\bra{q_i} w_\tau(0) \ket{q_i'} \,
\tilde{J}(q_f,q_f,t;q_i,q_i',0)
\label{c_pm11} \\
&&\tilde{J}(q_f,q_f,t;q_i,q_i',0) = 
\int\! {  D} q {  D} q' 
\exp \left [ \,  i ({  S}^{}_{-} [q]- {  S}^{}_{+} [q']) \, \right ] \
{\mathcal F}_{FV}[q,q']
\label{c_pm2}
\end{eqnarray}
with the path integration 
constraints
$q(0) = q_i$, $q(t)= q_f$,
$q'(0) = q'_i$, $q'(t)= q_f\,$.
The influence of the harmonic modes enters 
the generalized ``propagating function'' 
$\tilde{J}(q_f,q_f,t;q_i,q_i',0)$ via the 
Feynman-Vernon influence functional 
\begin{eqnarray}
{\mathcal F}_{FV}[\eta,\xi] &=& \exp \biggl\{  
\int_0^t  dt'  \int_0^{t'}  dt''  \;
\bigl[ {\dot \xi}(t')Q'(t'-t''){\dot \xi}(t'') 
\nonumber\\ && \hskip120pt
+ \, 
i  {\dot \xi}(t')Q''(t'-t'') {\dot\eta}(t'') \bigr]
\biggr\} \; ,
\label{FV}
\end{eqnarray}
where 
we have introduced the linear combinations
$$\eta(t)=\big[\,q(t)+q'(t)\,\big]/2\;,\qquad \qquad 
\xi(t)=\big[\,q(t)-q'(t)\,\big]/2 \;,$$
describing respectively propagation along the diagonal of the density matrix 
and off-diagonal excursions.
For a spin these paths are piecewise constant, the time intervals in which
the spin is in a diagonal state $\eta = \pm 1$ are called sojourns, while the 
time intervals
spent in an off-diagonal state $\xi =\pm 1$ are referred to as blips.
Depending on the initial condition for the spin, the initial state can
be either a sojourn or a blip. 
We now choose for the spin a diagonal initial state in the ``coordinate'', i.e.
$\tau_z$, basis $w_\tau(0) = \frac{1}{2} \, \hat I \, + 
\,\frac{1}{2} \delta p(0) \,\tau_z$,
where $\delta p(0)$ denotes the initial population difference between the 
$\tau_z$ eigenstates.
Then all paths that contribute to $C_{-+}(t)$ in Eq.(\ref{c_pm11})
start out and end up in a  sojourn state, 
$q(0)=q'(0)=q_i \rightarrow \eta(0)=\eta \equiv q_i$,
$q(t)=q'(t)=q_f \rightarrow \eta(t)=\eta' \equiv q_f$.
Thus a general path is characterized by $2m$ transitions at the intermediate 
times $t_j$ $(t_j =1, \, 2, \dots, 2m)$ and it is parametrized by
\begin{equation}\begin{array}{rcl}
 \xi^{(2m)}(t')&=&{\displaystyle \sum_{j=1}^m\xi_j [\Theta(t'-t_{2j-1})-
\Theta(t'-t_{2j})]\;,} \\
\eta^{(2m)}(t')&=&{\displaystyle \sum_{j=0}^{m}\eta_j [\Theta(t'-t_{2j})-
\Theta(t'-t_{2j+1}) ]\;} \, ,
\end{array} \label{path1}
\end{equation}
where $t_0=0$ and $t_{2m+1}=t$.
 Inserting the path (\ref{path1}) into $S_{\pm}$ in Eq.(\ref{c_pm2}) we get
\begin{equation}
\exp  \Bigl[ \, i ( {  S}^{}_{-} [q] -  {  S}^{}_{+} [q'] ) \, \Bigr]
= 
 \eta \eta'
 \left ( - \frac{\Delta}{2} \right )^{2m} \, B_m \,  
\exp \Bigl[ i v \sum_{j=0}^{m} \eta_j s_j \Bigr] \,
\label{newact}
\end{equation}
where $B_m=\exp \left [ -i \epsilon_{SB} \sum_{j=1}^{m} \xi_j \tau_j \right ]$
is the usual bias factor which depends only on the blip lengths
$\tau_j$. 
Notice that due to the two propagators $K_{\pm}$ which occur in 
Eq.(\ref{c_pm1}),
we get {\em dynamic} contributions which depend on the {\em sojourn}
lengths $s_j$ in the spin actions (last term in 
Eq.(\ref{newact})). \\
Substituting the paths (\ref{path1}) into Eq.(\ref{FV}),
the  influence functional reduces to    
\begin{equation}
\label{Ffv}
{\mathcal F}_{m} = G_{m}H_{m}\;.
\end{equation}
The factor  $G_{m}$ contains all the inter- and intra-blip-interactions, 
$$
\begin{array}{rcl}
  G_{m} &=& \exp \big[ - \sum_{j=1}^{m} Q'_{2j, 2j-1} -
                  \sum_{j=2}^{m} \sum_{k=1}^{j-1}
                  \xi_j \xi_k \Lambda_{j,k} \big]\;, 
\\
\Lambda_{j,k} &=& Q{}'_{2j,2k-1} + Q{}'_{2j-1,2k} - Q{}'_{2j,2k} - 
Q{}'_{2j-1,2k-1} \;,
\end{array}
$$
the  sojourn-blip interactions are captured by the phase factor $H_{m}$ 
given by
$$
\begin{array}{rcl}
H_{m}&=& {\displaystyle
\exp \big[ i \,
        \sum_{j=1}^m \sum_{k=0}^{j-1} \,\xi_{j}^{} \eta_{k}^{}\,X_{j,k}\;, 
        \big] }\\
X_{j,k}&=& {\displaystyle Q^{''}_{2j, 2k+1} + Q^{''}_{2j-1, 2k}- 
Q^{''}_{2j, 2k} - Q^{''}_{2j-1, 2k+1}}\;.
\end{array}
$$

The kernels $Q'$ and $Q''$ are the real and imaginary parts of 
the second integral of the bath correlation function
$$
 Q(t) \;=\; \int_0^\infty \! d\omega \;
\frac{G(\omega)}{\omega^2}  
\; \Bigl\{  \coth \Bigl(\frac{ \omega\beta}{2} \Bigr) \, 
\Bigl( 1-\cos(\omega t) \Bigr)
+\; i \sin(\omega t) 
\Bigr\} \;.
$$
and we have used the notation $Q_{h,i}=Q(t_h-t_i)$. \\
We proceed considering the important case of an ohmic bath 
Eq.~(\ref{eq:powerspec-bosons}), 
where the correlator $Q(t)$ in the scaling limit 
$\omega_c t \gg 1$, reads 
\begin{equation}
Q(t)=  2K \ln\left(\frac{ \beta\omega_c}{\pi}\,
\sinh\frac{\pi |t|}{ \beta}\right) + i \, \pi K {\rm sgn}(t)
\label{Qs} \; .
\end{equation}
Using the above scaling form each sojourn interacts only with the 
subsequent blip, so that $H_{m}$ simplifies to
\begin{equation}
H_{m} = \exp \Big[ i \pi K \sum_{k=0}^{m-1} \eta_k \xi_{k+1} \Big]\, .
\label{b-sS}
\end{equation}
Taking into account this features,
the sum over histories of paths of Eq.(\ref{c_pm2}) is represented by
\begin{description}
\item[$(i)$] the sum over all possible numbers of steps,
\item[$(ii)$] the time-ordered integrations over the  corresponding flip times
        $\{t_j\}$,
\item[$(iii)$] the sum over all possible arrangements of  blips 
        $\{\xi_j =\pm 1\}$ and sojourns $\{\eta_j = \pm 1\}$. 
\end{description}
The correlator $C_{-+}(t)$ becomes 
\begin{eqnarray}
\label{offd}
 C_{-+}(t) &&= 
\sum_{\eta} \, w_{\eta} \, e^{i v\eta t }  +  
\sum_{\eta} \, \eta \, w_{\eta} \sum_{\eta'} \, \eta' \times
\, \nonumber \\ &&
\sum_{m=1}^{\infty} 
                 \left(- \frac{\Delta}{2}\right )^{2m}
          \int_{0}^{t}\!{\mathcal D}_{2m}\{t_j\}
            \sum_{\{\xi_j\}} G_{m} B_{m}  
\sum_{\{\eta_j\}'} H_{m} 
\exp 
\Big[ i v \sum_{j=0}^{m} \eta_j s_j 
\Big] \, , 
\end{eqnarray}
where $w_\eta \equiv \bra {\eta} w_\tau(0) \ket {\eta}$ and
$\sum_{\{\eta_j\}'}$ denotes the sum over all paths $\eta_j$ subject
to the constraints $\eta_0=\eta$, $\eta_m=\eta'$.
The first term in Eq.(\ref{offd}) comes from the path without any
transition between the initial and final sojourns, and occurs for $\eta=\eta'$.
The sum over $\{\eta_j\}'$ can be easily performed using the 
scaling form (\ref{Qs}) for $H_m$, this leads to
\begin{eqnarray}
&&\sum_{\{\eta_j\}'} H_{m} 
\exp \Big[ i v \sum_{j=0}^{m} \eta_j s_j \Big] 
= 
e^{i \pi K \eta \xi_1} \, e^{i v(\eta s_0 + \eta' s_m)} 
\, 2^{m-1} \, D_m
\label{hm}
\end{eqnarray}
where $D_m=\prod_{j=2}^{m} \cos \left (\pi K \xi_{j}+ v s_{j-1} \right ) $.
Whereas for $v=0$ 
the sum over the intermediate sojourns provides a 
contribution which only depends on the blip states $\xi_j$,
because of the coupling the sum over $\eta_j$ gives also a {\em dynamic}
term dependent on the sojourn lengths $s_j$.
Thus we finally get
\begin{eqnarray}
&& C_{-+}(t) =
\sum_{\eta} \, w_{\eta} \, e^{i v\eta t }  +  
\tilde{A}(t) + i  \tilde{B}(t) 
\label{formal}
\\ 
&&\tilde{A}(t)= \frac{1}{2}\sum_{\eta} \, \eta w_\eta 
\sum_{\eta'} \,  \eta' A(\eta, \eta', t) 
\label{Adef}
\\
&&\tilde{B}(t)= \frac{1}{2}\sum_{\eta} \,  w_\eta 
\sum_{\eta'} \,  \eta'  \, B(\eta, \eta', t)
\label{Bdef}
\end{eqnarray}
where
\begin{eqnarray}
&&\!\!\!\!\!\!\!\!\!\!\!\!\!\!\! A(\eta, \eta', t) =
\sum_{m=1}^{\infty} \left(- \frac{\Delta^2}{2}\right )^{m} 
\int_{0}^{t}\!{\mathcal D}_{2m}\{t_j\}
e^{i v(\eta s_0 + \eta' s_m)} 
\sum_{\{\xi_j\}} G_{m} B_{m} D_m  \cos(\pi K) 
\label{A} \\
&&\!\!\!\!\!\!\!\!\!\!\!\!\!\!\!
 B(\eta, \eta', t) = 
\sum_{m=1}^{\infty} \left(- \frac{\Delta^2}{2}\right )^{m} 
\int_{0}^{t}\!{\mathcal D}_{2m}\{t_j\} 
e^{i v(\eta s_0 + \eta' s_m)} 
\sum_{\{\xi_j\}} \xi_1 G_{m} B_{m} D_m  \sin(\pi K) 
\, .  
\label{B}
\end{eqnarray}
Eq.(\ref{element}) with Eqs.(\ref{formal}) - (\ref{B})  
is the exact formal expression for
the coherence  $\rho_{01}(t)$, valid for Ohmic
damping in the scaling regime.

\section{Generalized Master Equation}
\label{sec:gen-master}
For quantum dissipative systems it is generally useful to formulate the
dynamics of the open system by means of appropriate
master equations 
for the reduced density matrix.
For the spin-boson problem generalized master equations relating the
averages of the spin components $\tau_\alpha$ have been obtained in 
Refs.~\cite{book,GME}.
In this section we show that the series expressions for $A(\eta, \eta',t)$ and
$B(\eta, \eta',t)$ given in Eqs.(\ref{A},\ref{B})
can also be cast into the form of exact integro-differential equations.

To this end we have to find the kernels, i.e. the irreducible components
in the formal expressions (\ref{A},\ref{B}).
Irreducibility of a kernel means that it cannot be cut into uncorrelated
pieces at an intermediate sojourn without removing correlations due to 
the harmonic bath, across this sojourn.  
Following the lines of Refs.\cite{book,GME} we define irreducible 
influence functions subtracting from 
the inter- and intra-blip interaction factors $G_m$ all the reducible 
components 
\begin{eqnarray}
\label{irreducible}
\tilde{G}_n &\equiv& G_n 
\! - \!\sum_{j=2}^{n} (-1)^{j} \!\!\!  \sum_{m_1 \dots m_j}
G_{m_1} 
\dots G_{m_j} 
\delta_{m_1+m_2+ \dots + m_j, n} \, , \nonumber
\end{eqnarray}
where the inner sum is over all positive integers $m_j$.
By inspection of (\ref{A},\ref{B}) we find that the 
irreducible kernels are
\begin{eqnarray}
K_1(t-t')& = & {\mathcal K}_1(t-t') \nonumber \\
&-&  
\cos(\pi K) \sum_{n=2}^{\infty} 
        \left( -{\Delta^2 \over 2}\right)^{n} \int_{t'}^{t} dt_{2n-1} ..
        \int_{t'}^{t_3} dt_{2} 
\sum_{\xi_j}  \tilde{G}_n B_n  
D_n
\label{ex-kerns} \\
K_2(t-t') &=& {\mathcal K}_2(t-t') \nonumber \\
&+&   \sin(\pi K) \sum_{n=2}^{\infty}  
        \left(- {\Delta^2 \over 2}\right)^{n} \int_{t'}^{t} dt_{2n-1} \dots 
        \int_{t'}^{t_3} dt_{2} 
\sum_{\xi_j}\xi_1 \tilde{G}_n B_n  D_n \, ,
\label{ex-kerna}
\end{eqnarray}
where the lowest order contributions do not depend on the coupling $v$
and coincide with the lowest order kernels entering the
spin dynamics in the uncoupled case~\cite{book,GME}
\begin{eqnarray}
\label{NIBA1}
{\mathcal K}_1(t-t') &=& 
\Delta^2 \cos(\pi K) G_1(t-t')\cos[\epsilon_{SB} (t-t')] \\
\label{NIBA2}
{\mathcal K}_2(t-t') &=&
\Delta^2 \sin(\pi K) G_1(t-t') \sin[\epsilon_{SB} (t-t')] \, .
\end{eqnarray}
With the above the generalized master equations for  $A(\eta, \eta',t)$ and 
$B(\eta, \eta',t)$ read
\begin{eqnarray}
\frac{d}{dt} A &=&  i \eta' v A 
- \int_{0}^{t} dt'\left[  e^{i \eta vt'} + A \right]K_1(t-t')
\nonumber \\
 &+& i v  \int_{0}^{t} dt'  e^{- i \eta' v(t-t')}
\int_{0}^{t'} dt''[ \, \eta'  K_1(t'-t'') 
-K_2(t'-t'') \,]  A(t'')  
\label{eqA2} \\
\frac{d}{dt} B &=& i \eta' v  B 
- \int_{0}^{t} dt' [ K_1(t-t') B 
- i K_2(t-t') e^{i \eta vt'} ]
 \nonumber \\
&+& i v  \int_{0}^{t} dt'  e^{- i \eta' v(t-t')} 
\int_{0}^{t'} dt''  [\,  \eta'K_1(t'-t'') 
- K_2(t'-t'') \, ]  B(t'')  \, .
\label{eqB2}
\end{eqnarray}
In the absence of coupling, $v=0$, $A(\eta, \eta',t)$ and 
$B(\eta, \eta',t)$ are simply related to the 
components of the average $\langle \tau_z \rangle_t$ symmetric and 
antisymmetric under bias inversion $\epsilon_{SB} \to -\epsilon_{SB}$,
$\langle \tau_z \rangle_t= P_s(t)+P_a(t)$, via
$A_{0}(\eta, \eta',t)=P_s(t)-1$ and
$B_0(\eta, \eta',t)= -i P_a(t)$.
In this limit
Eqs.(\ref{eqA2},\ref{eqB2}),
reduce to the master equations satisfied by $P_s(t)$ and 
$P_a(t)$~\cite{book,GME}.

Being of convolutive forms the integro-differential equations
Eqs.(\ref{eqA2},\ref{eqB2}) are conveniently solved in Laplace space.
Performing the sums in Eqs.(\ref{Adef},\ref{Bdef})
the Laplace transform of the correlator $C_{-+}(t)$ reads
\begin{eqnarray}
{\widehat C_{-+}(\lambda)}& = & \frac{1}{D(\lambda)} \, \left [ \, 
\lambda + K_1(\lambda) 
- i v \delta p(0) \, \right ] \\
D(\lambda) &=& \lambda^2 +  v^2 
+ \lambda K_1(\lambda) +
i v K_2(\lambda) \, ,
\label{eq:poles-exact}
\end{eqnarray}  
where $K_1(\lambda)$, $K_2(\lambda)$ are the Laplace transforms of the
kernels (\ref{ex-kerns},\ref{ex-kerna}) and
$\delta p(0)$ comes from the initial condition chosen for the
spin $\vec{\tau}$.

\section{Reduced dynamics at finite temperatures}
\label{sec:reduced-dynamics}
Solving the pole equation (\ref{eq:poles-exact}) represents a formidable task,
so in the following we shall focus on special damping and temperature
regimes which are particularly relevant for quantum computing. 
We shall analyze dephasing at  finite temperatures at two special coupling
conditions: weak damping $K \ll 1$ and $K=1/2$.
In the second case the spin-boson model can be mapped onto a Fano Anderson 
model~\cite{mahan} which is suitable to study 
dephasing due to  bistable fluctuators producing $1/f$ noise~\cite{PRL}. 

We shall adopt an approximate treatment  widely
used in the spin-boson literature~\cite{book,Leggett},
the noninteracting-blip approximation (NIBA).
In the NIBA, the blip-blip interactions $\Lambda_{j,k}$ are neglected,
and also the sojourn-blip interactions $X_{j,k}$ are disregarded except 
those of neighbors, $k=j-1$, and they are approximated by 
$X_{j,j-1} = Q{''}(\tau_j)$.
Thus the influence functions (\ref{Ffv})
factorize into the individual blip influence factors,
and the irreducible influence functions (\ref{irreducible}) vanish
for $n >1$. 
For Ohmic damping in the scaling regime, 
each blip interacts with the subsequent sojourn  only (see Eq.(\ref{b-sS})), 
thus the only approximation involved in the
NIBA is  the neglect of the  blip-blip interactions. 
This approximation is justified in the regimes of our interest, 
see ref.~\cite{book}.

Within the NIBA the series expression for the irreducible kernels
(\ref{ex-kerns},\ref{ex-kerna})
reduce to the lowest order contributions (\ref{NIBA1},\ref{NIBA2}).
Evaluation of Eqs.(\ref{eqA2},\ref{eqB2}), with these kernels
give the reduced dynamics of the qubit in the NIBA.
To make things simpler we shall consider the unbiased case $\epsilon_{SB}=0$. 

\subsection{The case $K=1/2$}
\label{sec:keyonehalf}
For  $K=1/2$ the Laplace transform of the 
kernel ${\mathcal K}_1(t-t')$ takes the form
$$
\lim_{K\rightarrow  1/2}
\int_0^{\infty} d\tau \,e^{- \lambda \tau}  {\mathcal K}_1(\tau)
 \, 
=\frac{\pi}{2}\frac{\Delta^2}{\omega_c} \equiv \gamma_F  \;,
$$
the zero coming from
$\cos(\pi K)$ for $K \to 1/2$ being compensated from
the short distance singularity of the intra-blip interaction
$\lim_{\tau \to 0} e^{-Q'(t)} \approx (\omega_c t)^{2K}$.
The pole equation (\ref{eq:poles-exact}) is readily solved and
$C_{-+}(t)$ reads
\begin{eqnarray} 
 && C_{-+}(t) =
A \, e^{-{1 \over 2} (1 - \alpha) \, \gamma_F t}
        \, + \,(1-A) \, e^{-{1 \over 2} (1 + \alpha) \, \gamma_F t} \, .
        \label{offdiag} 
\end{eqnarray}
This form is suitable for illustrative purposes 
since $\alpha$ and $A$ have a very simple expression~\cite{DPG} 
\begin{equation}
\alpha = 
\sqrt{1-  \Bigl({2 v \over \gamma_F}\Bigr)^2 }
\qquad ; \qquad 
A = {1 \over 2 \alpha} \, 
\Bigl[ 1+ \alpha - i  \, 
{2 v \over \gamma_F} \, \delta p(0)\Bigr]\, \, .
\label{alpha} 
\end{equation}
The behavior of $C_{-+}(t)$ has been analyzed in Ref.~\cite{PRL} 
where two qualitatively different regimes for the problem were identified.

\vspace{0mm}\noindent
{\em Fast environment} --\hspace{2mm} If $2 v/\gamma_F \ll 1$ 
only one of the exponentials
appearing in Eq.~(\ref{offdiag})
dominates the long-time behavior and $C_{-+}(t)$ shows a single rate, 
given by ${1 \over 2} (1 - \alpha) \, \gamma_F \approx v^2/\gamma_F$,
the standard Golden Rule result. The 
decay of $C_{-+}(t)$ at all times, described by 
$\Gamma(t)= - \ln \{ C_{-+}(t)\}$, is well approximated by simulating the 
SB model with suitable set of harmonic oscillators reproducing the
power spectrum $S_\tau(\omega)$. In this case the decay is described by
$\Gamma(t) \approx \Gamma_{osc}(t)$ 
where~\cite{kn:palma96,book}  
\begin{equation}
\label{kn:gamma-osc}
\Gamma_{osc}(t)
\,=\, 
\int_0^{\infty} {d \omega \over \pi} \; {S}_\tau(\omega)
\; {1 - \cos \omega t \over \omega^2} \, .
\end{equation}
Here  the power spectrum Eq.(\ref{eq:powerspec}) is a  
Lorentzian $S_\tau(\omega) =  2 v^2 \gamma_F/(\gamma_F^2 + \omega^2)$,
due to the pure relaxation behavior of $\langle \tau_z \rangle$ for
$v=0$~\cite{book}.
Thus in this regime the spin $\vec{\sigma}$ sees the SB environment as an 
effective set of harmonic oscillators, and no detailed information or memory
effects of the environment enters $C_{-+}(t)$.

\begin{figure}[h!]
\label{fig} 
\resizebox{65mm}{50mm}{\includegraphics{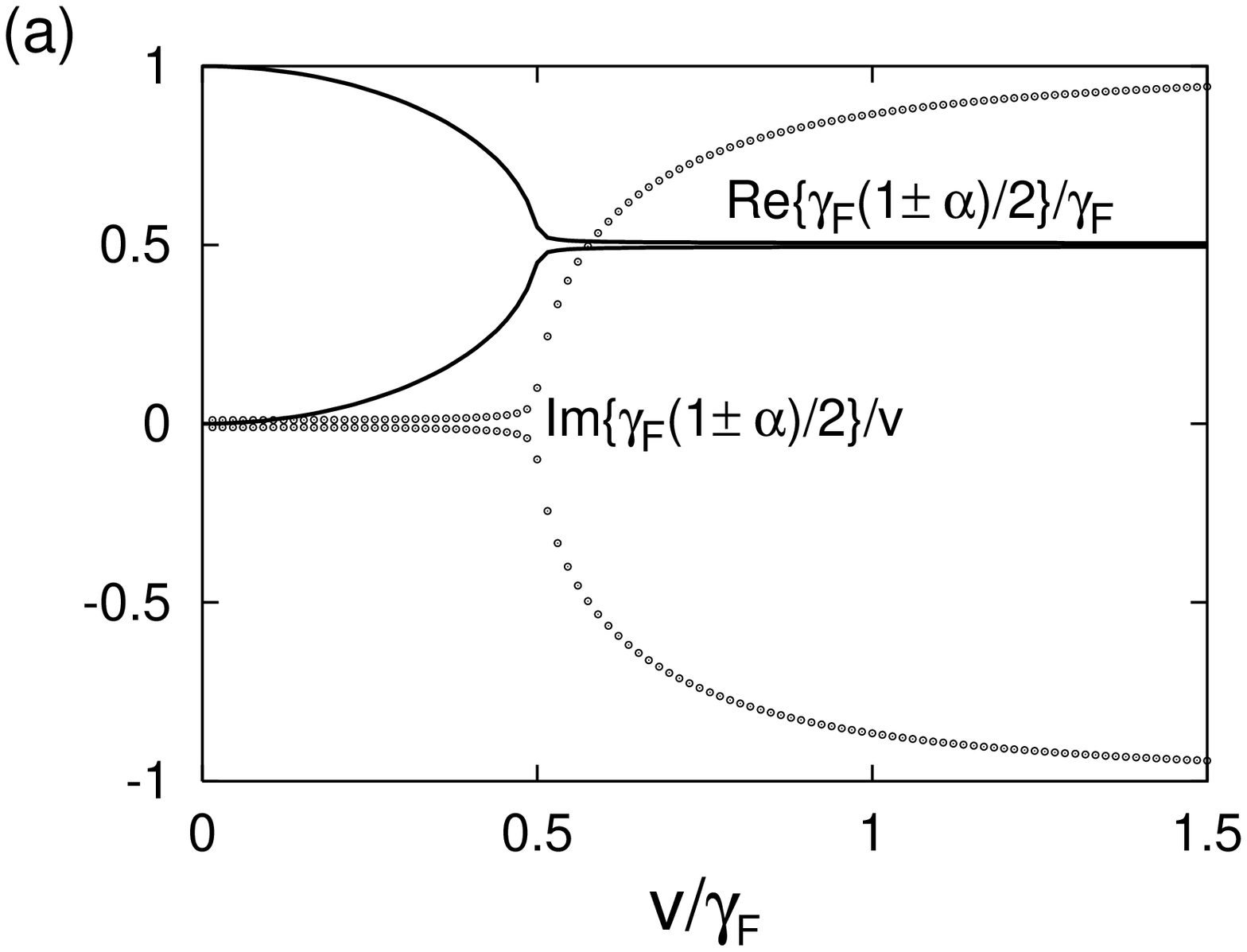}}
\hfill
\resizebox{70mm}{!}{\includegraphics{long-times.eps}}
\caption{(a) The real part (solid lines) and the imaginary part (dots) 
of the rates entering 
Eq.~(\ref{offdiag}) are plotted for finite temperature 
$k_B T \gg \epsilon_{SB}>0$. (b) The full expression for $\Gamma(t)$ 
for various $2v/\gamma_F$: deviations from the weak coupling result 
$\Gamma_{osc}(t)$ (dashed line) are apparent for $\gamma_F < 2v$. Here 
$\delta p = \tgh[\epsilon_{SB}/(2k_B T)]=0.08$.}
\end{figure}

\vspace{0mm}
{\em Slow environment} --\hspace{2mm} If $2v/\gamma_F \gg 1$, 
$C_{-+}(t)$ is determined by both the exponential terms 
in Eq.~(\ref{offdiag}). Deviations from the weak coupling result
$\Gamma_{osc}(t)$  are observed, in particular the 
short times behavior $t \ll \gamma_F^{-1}$ shows
memory effects whereas for long
times the  coupling $v$ does not enter the 
decay rates, given by $\gamma_\infty = \gamma_F/2$, but determines 
the frequency shifts, which reduce to $\pm v$ (see Fig.~\ref{fig}.a).

We mention that for $K=1/2$ the same structure Eq.~(\ref{offdiag}) is also 
found in more general situations, $\epsilon_{SB}, E_J \neq 0$~\cite{PRL,DPG}. 
Formulas for $A$ and $\alpha$ are more complicated, for instance 
$\MyRe \alpha$ and $\MyIm \alpha$ 
are both non vanishing in the entire $2 v/\gamma_F$ range and a third regime 
$2 v/\gamma_F \sim 1$ appears (see Fig.\ref{fig}).  
However the analysis using the expressions (\ref{alpha}) 
remains still valid, and in the following we will identify the different 
regimes for $K\ll 1$ 
from a similar analysis of the poles of ${\widehat C_{-+}(\lambda)}$.

\subsection{Weak ohmic damping $K \ll 1$ and finite temperatures}

For weak ohmic damping, $K \ll 1$ and temperatures
$k_B T_b \leq  k_B T \ll \omega_c$, where
$k_B T_b = \sqrt{\Delta{}_r^2+v^2}\,$,
the characteristic fluctuations of the harmonic
oscillators in $H_{SB}$ have no memory and $Q(t)$ assumes the Markov form
$$Q(t) = 2K \Big\{ {\pi |t| \over \beta} + 
\ln{\Big( {\beta \omega_c \over 2 \pi}\Big)} \Big\} +i \pi K {\rm sgn}(t)\, .
$$
Using this latter expression the interblip interactions cancel out exactly, 
thus justifying the NIBA. 

In the absence of bias, $\epsilon_{SB}=0$ only the kernel 
${\mathcal K}_1(\lambda)$ is non vanishing, 
and in the Markov limit it reads 
${\mathcal K}_1(\lambda) = \Delta_T^2 /[ \lambda+2 \gamma(T)]$, 
where $\gamma(T)$ is given by Eq.~(\ref{eq:eff-coupling}) and
we have introduced the temperature dependent tunneling term
$\Delta_T = \Delta_r (2 \pi k_B T/ \Delta_r)^K$.
The correlator ${\widehat C_{-+}(\lambda)}$ is readily found as 
\begin{eqnarray}
{\widehat C_{-+}(\lambda)}& = & {[\lambda+2\gamma(T)] \,
[\lambda - i v \delta p(0)] + \Delta_T^2 \over N(\lambda)} 
\label{eq:C-laplace}
\\
N(\lambda) &=& 
(\lambda^2 +  v^2) \, [\lambda+2\gamma(T)] + \Delta_T^2 \lambda 
\nonumber 
\end{eqnarray}
The qualitative behavior of $C_{-+}(t)$ is determined by the 
three poles found by
solving the equation $N(\lambda) = 0$.

\vspace{2mm}\noindent
{\em Low temperatures} --\hspace{2mm} 
At temperatures
$T_b \le T \ll T_0$
the poles, in lowest order in $T/T_0$, are given  by
\begin{eqnarray}
\label{lowTl0}
\lambda_0 \,=\, - \frac{2 v^2 \,\gamma(T) }{v^2+\Delta{}_T^2} 
\qquad ; \qquad 
\lambda_{1/2} \,=\, - \frac{\Delta{}_T^2 \,\gamma(T) }{v^2+\Delta{}_T^2}
\pm i \sqrt{v^2+\Delta{}_T^2} \, .
\end{eqnarray}
The limiting temperature is given by
$k_B T_0 \approx  \Delta_T/(\pi K) - v^2/(2 \pi K \Delta_T)$ 
for $v/\Delta_T \ll 1$ and by $k_B T_0 = \sqrt{\Delta_T^2+v^2}/(2 \pi K)$ for
$v/\Delta_T \gg 1$. 
The behavior of $C_{-+}(t)$ depends essentially on 
$v/ \Delta_T$. If $v / \Delta_T \ll 1$ the reduced dynamics 
shows a single rate, 
corresponding to the dominant pole 
$\lambda_0 \approx - 2 (v/ \Delta_T)^2 \, \gamma(T)$.
Instead in the opposite regime, $v / \Delta_T \gg 1$, 
the two poles $\lambda_{1/2} \approx -(\Delta_T/v)^2  \gamma(T)
\pm i v$ dominate.  
The frequency shifts are given by $\pm v$ as in the slow environment
case of section~\ref{sec:keyonehalf}. Thus for $v/ \Delta_T \sim 1$ a 
crossover is expected in the behavior of  $C_{-+}(t)$.

\vspace{2mm}\noindent
{\em High temperatures} --\hspace{2mm} 
If $T \gg T_0$, the leading behavior is given by
\begin{eqnarray}
\lambda_0 \,=\, -  2 \gamma(T) 
\qquad ; \qquad
\lambda_{1/2} \,=\, -\frac{\Delta{}_T^2}{4 \gamma(T)} \pm \frac{1}{2}  
\sqrt{ \frac{\Delta{}_T^4}{4 \gamma^2(T) } - 4 v^2} \, .
\end{eqnarray}
In this case there is an intermediate temperature regime,
$k_BT_0 \ll k_B T < k_B T_+ = \Delta_T^2/(4 \pi K v)$,  
where we may have three real roots. In this regime, which occurs only if
$v / \Delta_T < 1$, the dominant contribution to
$C_{-+}(t)$ comes again from a single pole, 
$\lambda_1 \approx - 2 (v/ \Delta_T)^2 \, \gamma(T)$. Instead for 
temperatures $T > T_+$ the two complex conjugate $\lambda_{1/2}$
are the dominant poles. 
The decay rate
displays the Kondo behavior 
$\Delta_T^2/[4 \gamma(T)] \propto T^{2K-1}$ and does not depend on $v$.
The coupling $v$ enters
the frequency shifts, which approach $\pm v$ if the temperature is 
further increased. Thus in the region $T \gg T_0$ and $v < \Delta_T$
we find a crossover which may be triggered by 
both $v$ and $T$, the crossover line being given by 
\begin{equation}
v \sim  {\Delta_T^2 \over 4\pi K \, k_B T} \;.
\label{eq:crossover-line}
\end{equation} 
Finally for larger values of $v > \Delta_T$ the system 
stays in the regime where the dominant rates are the two complex conjugate
$\lambda_{1/2} \approx  - \Delta{}_T^2 / [4 \gamma(T)]  \pm i v$.
Again $v$ does not enter the decay rate but determines the shifts 
$ \sim \pm v$. 

\section{Discussion of the results and conclusions}
\label{sec:conclusions}
The analysis of the poles of ${\widehat C_{-+}(\lambda)}$, 
Eq.(\ref{eq:C-laplace}) allows to extend the qualitative 
conclusion for $K=1/2$ to the regime $K \ll 1$. 
In this latter case we find that the dynamics of 
$\vec{\sigma}$ shows a behavior which reflects
the crossover of the SB environment to a slow dynamics regime.
For $K \ll 1$ the SB environment can be modulated by acting on both 
the coupling $v$ and the temperature. Instead for 
$K=1/2$ and $\epsilon_{SB}=0$ 
the crossover is triggered by the parameter
$v/\gamma_F$ only. 
This is simply due to the fact that
for the relatively strong coupling $K=1/2$, $\tau_z$ has a relaxation
dynamics at any temperature. Thus at any temperature we may identify  
the correlation time of
the SB environment $\tau_c \approx \gamma_F^{-1}$. The 
crossover to the slow environment regime occurs when $2 v \tau_c \sim 1$ which 
marks the limit where the standard master equation approach in the coupling
$v$ would break down~\cite{kn:cohen}.

To understand the behavior for $K \ll 1$ we focus on $v \ll \Delta_T$ 
and we identify the correlation time 
$\tau_c$ of the SB environment from the dynamics of the SB model for $v=0$. 
In this case
$\tau_z$ displays a crossover from coherent to
incoherent dynamics at a temperature 
$k_B T^*(K) \approx \Delta_r / \pi K$.
Below $T^*$ coherent oscillations at frequency $\approx \Delta_T$
are slowly damped with rate given by $\gamma(T) \ll \Delta_T$ and 
we may identify $\tau_c \sim \Delta_T^{-1}$. Thus the SB environment is
fast, since $2 v \tau_c \approx 2 v / \Delta_T \ll 1$ and
$\rho_{01}(t)$ shows the single rate 
$\lambda_0 \approx - 2 (v/ \Delta_T)^2 \, \gamma(T) \approx S_\tau(0)/2$.
If the temperature increases fairly above $T^*$ 
the dynamics of $\vec{\tau}$ is incoherent and 
relaxation occurs, with a slow rate
$\gamma_r(T) \approx \Delta_T^2/2\gamma(T)$~\cite{book} which is
identified with $\tau_c^{-1}$. Slow environment effects are then expected to 
appear in the reduced dynamics of $\vec{\sigma}$ at 
$2 v / \gamma_r(T) \sim 1 $ which is indeed the crossover line 
Eq.(\ref{eq:crossover-line}).

Finally we point out possible extensions of our analysis. 
A SB environment with subohmic (superohmic) spectral densities~\cite{book}
could be treated along the same lines, but since the uncoupled SB dynamics
is always slow (fast) we do not expect tunable crossover lines to appear.
Our analysis can be easily extended to an 
environment made of an arbitrary number of SB impurities. In this case
$C_{-+}(t) = \prod_i C^{(i)}_{-+}(t)$ where 
$ C^{(i)}_{-+}(t)$ is the individual contribution of the $i-$th SB impurity.
This allows to model a simple environment which produces $1/f$ noise by 
choosing a suitable distribution~\cite{kn:weissman} 
of the parameters of the set SB impurities. 
For instance in Ref.~\cite{PRL}, which focuses on $K=1/2$, 
the standard distribution $\propto 1/\gamma_F$ has been chosen for the 
relaxation rate of the SB. 
We notice that for $K=1/2$ temperature has no effect in the regime we consider.
If we allow for impurities with $\varepsilon_{SB} \neq 0$ the temperature 
determines at most how many impurities in the set are active. 
Instead for a $1/f$ environment modeled by a set of 
ohmic SB impurities with $K \ll 1$ our results show a 
non trivial behavior as a function of the temperature, which may be relevant 
for recent observations~\cite{kn:vanharlingen} of temperature-dependent 
dephasing due to $1/f$ noise.
Finally if $E_J \neq 0$, $\sigma_z$ is not conserved 
and relaxation occurs. The relaxation rate increases monotonically 
with  $E_J/\varepsilon$. We believe that the physical picture emerging from
this work  still applies to dephasing (see Ref.~\cite{DPG} for $K=1/2$) 
at least for times smaller than the relaxation time.

\section{Acknowledgements}
The authors are pleased to thank Uli Weiss for the physics they learned
and enjoyed from the extremely vivid cooperation with him during the
last 15 years.


\begin{thebibliography}{00}
\bibitem{kn:qcomp} 
        {A. Ekert and A. Jozsa}, {Rev. Mod. Phys.} {\bf 68}, {733} (1996);
        {\em Quantum Computation and Quantum Information Theory},
        edited by {C. Macchiavello, G.M. Palma, A. Zeilinger},  
        World Scientific (2000).
        {M. Nielsen and I. Chuang}
        {\em Quantum Computation and Quantum Communication},
        Cambridge University Press, (2000).
        {\em Experimental implementation of quantum computation},
        edited by {R.G. Clark}, Rinton Press, Princeton (2001). 
\bibitem{kn:zurek}
        {W.\ Zurek}, {Physics Today} {\bf 44}, {36} (1991).
\bibitem{kn:palma96}
        {G.M. Palma, K.-A. Suominen and A.K. Ekert}, 
        {Proc. Roy. Soc. London A} {\bf 452}, {567} (1996).
\bibitem{kn:rmp} 
        {Y. Makhlin, G. Sch\"on and A. Shnirman}, 
        {Rev. Mod. Phys.} {\bf 73}{357} (2001).
\bibitem{book} {U. Weiss} {\em Quantum Dissipative Systems} 2nd Ed 
        (World Scientific, Singapore 1999).               
\bibitem{Leggett} {A. Leggett  {\em et al.}}, {Rev. Mod. Phys.} {\bf 59}, {1} (1987).
\bibitem{kn:dissipative} 
{A.O. Caldeira and A.J. Leggett},
         {Ann. Phys.} {\bf 149}, {374} (1983).
\bibitem{kn:cohen} 
        {C. Cohen-Tannoudji, J. Dupont-Roc and G. Grynberg}
        {\em Atom-Photon Interactions}, Wiley-Interscience (1993).
\bibitem{PRL}
        {E.\ Paladino {\em et. al.}}, {Phys. Rev. Lett.} {\bf 88}, {228304} (2002);
        G. Falci {\em et. al.}, Proceedings of the 
        International School Enrico Fermi on "Quantum Phenomena of Mesoscopic Systems", B. Altshuler and V. Tognetti Eds., IOS Bologna (2003).
\bibitem{kn:weissman} 
        {M.B.\ Weissman}, {Rev.\ Mod.\ Phys.} {\bf 60}, {537} (1988);
        {A.B.\ Zorin {\em et. al}}, {Phys.\ Rev.\ B} {\bf 53}, {13682} (1996).
\bibitem{kn:exp} {Y.\ Nakamura, Yu.A.\ Pashkin, J.S.\ Tsai}, 
        {Nature} {\bf 398}, {786} (1999); {Y. \ Nakamura {\em et. al.}}, 
        { Phys. Rev. Lett.} {\bf 88}, {047901} (2002); {D. Vion {\em et al.}}, 
        {Science} {\bf 296}, {886} (2002); 
        {Y. Yu \em et al.}, {Science} {\bf 296}, {889} (2002);
        {J. Martinis \em et al.}, {Phys. Rev. Lett.} {\bf 89}, {117901} (2002);
        {J. Friedman \em et al.}, {Nature} {\bf 406}, {43} (2000);
        {I. Chiorescu {\em et al.}} Science, {\bf 299}, 1869 (2003); 
        {Yu. A. Pashkin {\em et al.}} Nature {\bf 421}, 823  (2003).
\bibitem{mahan} Fano impurities are localized levels  
occupied by electrons which 
may tunnel to a band of delocalized states, see G. D. Mahan,
        {\em Many-particle physics}, third edition, Kluwer 2000.
      \bibitem{kn:josephson-qubit-theory} {D.A.\ Averin}, {Sol.\ State
          Comm.} {\bf 105}, {659} (1998); {J.E.\ Mooij {\em et al.}},
        {Science} {\bf 285}, {1036} (1999); {Y. Makhlin, G. Sch\"on
          and A. Shnirman}, {Nature} {\bf 398}, {305} (1999);
        {G. Falci \em et al.}, {Nature} {\bf 407}, {355} (2000).
\bibitem{kn:grifoni-governale} M. Governale, M. Grifoni, G. Sch\"on,
        Chem. Phys. {\bf 268}, 273 (2001).
\bibitem{kn:devoret-schoelkopf} 
        R. Aguado and L.P Kouwenhoven, Phys. Rev. Lett. {\bf 84}, 
        1986 (2000); 
        M. H. Devoret and R. J. Schoelkopf, Nature {\bf 406}, 1039 (2000).
\bibitem{GME} M. Grifoni, M. Sassetti and U. Weiss, Phys. Rev. E {\bf 53},
           R2033 (1996);
        M. Grifoni, E. Paladino, U. Weiss, 
        Eur. Phys. J. B {\bf 10}, 719 (1999).
\bibitem{DPG} 
E. Paladino, L. Faoro, G. Falci, Adv.  Solid State Phys. {\bf 43}, 747 (2003).
\bibitem{kn:vanharlingen} 
J. A. Bonetti, D. J. Van Harlingen, M. B. Weissman,
Physica C, {\bf 288 - 389}, 343 (2003).
\end{thebibliography}
\end{document}